\documentclass[twocolumn,amsmath,amssymb,aps,pra]{revtex4-2}

\usepackage{graphicx}
\usepackage{dcolumn}
\usepackage{bm}
\usepackage{color}

\begin{document}

\title{Double-Transmon Coupler: Fast Two-Qubit Gate with No Residual Coupling for Highly Detuned Superconducting Qubits}

\author{Hayato Goto}
\affiliation{
Frontier Research Laboratory, 
Corporate Research \& Development Center, 
Toshiba Corporation, 
1, Komukai Toshiba-cho, Saiwai-ku, Kawasaki-shi, 212-8582, Japan}

\date{\today}

\begin{abstract}

Although two-qubit entangling gates are necessary for universal quantum computing, 
they are notoriously difficult to implement with high fidelity. 
Recently, tunable couplers have become a key component for realizing high-fidelity two-qubit gates in superconducting quantum computers. 
However, it is still difficult to achieve tunable coupling free of unwanted residual coupling, in particular, for highly detuned qubits, 
which are desirable for mitigating qubit-frequency crowding or errors due to crosstalk between qubits. 
We thus propose this kind of tunable coupler, which we call a double-transmon coupler, 
because this is composed of two transmon qubits coupled through a common loop with an additional Josephson junction. 
Controlling the magnetic flux in the loop, we can achieve not only fast high-fidelity two-qubit gates, 
but also no residual coupling during idle time, where computational qubits are highly detuned fixed-frequency transmons. 
The proposed coupler is expected to offer an alternative approach to higher-performance superconducting quantum computers.

\end{abstract}

\maketitle

\section{Introduction}

Remarkable advances have been made 
in the technologies for realizing quantum computers over the past decade. 
Nevertheless, two-qubit entangling gates, which are necessary for 
universal quantum computing together 
with single-qubit gates~\cite{Deutsch1995a,Barenco1995a,DiVincenzo1995a,Sleator1995a,Lloyd1995a,Barenco1995b,Nielsen}, 
are still hard to implement with high fidelity. 
For instance, two-qubit gates with fidelity of over 99\%, 
which is necessary for tasks such as fault-tolerant quantum computation 
using quantum error correction~\cite{Raussendorf2007a,Raussendorf2007b,Fowler2009a,Wang2011a,Fowler2012a}, 
have been demonstrated experimentally by using only a few kinds of physical systems: 
laser-cooled trapped ions~\cite{Ballance2016a,Gaebler2016a,Schafer2018a,Pino2021a,Egan2021a,Ryan2021a}, 
superconducting circuits~\cite{Barends2014a,Kelly2015a,Arute2019a,Foxen2020a,Xu2020a,Wu2021a,Ficheux2021a,Negirneac2021a,Sung2021a,Stehlik2021a,Sete2021a,Kandala2021a,Mitchell2021a,Ye2021a}, 
and most recently silicon-based approaches such as quantum dots~\cite{Noiri2022a,Xue2022a,Mills2022a} and 
donor spins~\cite{Madzik2022a}. 
Among these, superconducting circuits may be promising in the sense that their qubits are, of course, 
solid-state devices and therefore do not need trapping, unlike trapped ions, 
and also two-dimensional qubit arrays have already been realized~\cite{Arute2019a,Wu2021a,Gong2021a}, 
which is still difficult for quantum dots, by recently developed 3D integration technologies~\cite{Yost2020a,Mallek2021a,Kosen2021a}.

Tunable couplers have recently become a key component for high-fidelity two-qubit gates 
in superconducting quantum computers~\cite{Arute2019a,Foxen2020a,Xu2020a,Wu2021a,Sung2021a,Stehlik2021a,
Sete2021a,Ye2021a,Mundada2019a,Li2020a,Collodo2020a,Ni2021a,Petrescu2021a,Jin2021a,Leroux2021a,Miyanaga2021a}. 
Tunable couplers allow us not only to implement fast two-qubit gates, 
but also to turn off an energy-exchange interaction called transverse or $XY$ coupling~\cite{Krantz2019a}. 
Major tunable couplers, including those used for demonstration of quantum supremacy (advantage)~\cite{Arute2019a,Wu2021a}, 
are based on the cancellation between a direct coupling via a capacitor 
and an indirect coupling via a frequency-tunable transmon qubit~\cite{Yan2018a}, 
which we refer to as a single-transmon coupler. 
This coupler is regarded as a capacitor or transmon version of a previously proposed inductor-based coupler 
for flux qubits~\cite{Niskanen2006a,Niskanen2007a}. 
(Other kinds of inductor-based tunable couplers have also been proposed~\cite{Allman2014a,Whittaker2014a,Chen2014a,Neill2018a}.) 
In the single-transmon coupler, 
there exists an unwanted correlated energy shift due to residual coupling 
called longitudinal or $ZZ$ coupling~\cite{Krantz2019a}. 
Residual $ZZ$ coupling has recently become a central issue 
in the field of superconducting quantum computers~\cite{Xu2020a,Sung2021a,Stehlik2021a,Sete2021a,Kandala2021a,Mitchell2021a,Ye2021a,
Mundada2019a,Li2020a,Collodo2020a,Ni2021a,Petrescu2021a,Jin2021a,Zhao2020a,Ku2020a,
Noguchi2020a,Xu2021a,Zhao2021a,Finck2021a,Tripathi2021a}. 
Some research groups have found special conditions under which the $ZZ$ coupling 
in the single-transmon coupler vanishes~\cite{Stehlik2021a,Li2020a}. 
However, the vanishing points exist only in a region of small detunings 
between two computational qubits~\cite{Stehlik2021a} (see Appendix~\ref{appendix-coupler}). 
In other words, in the single-transmon coupler, there is inevitable $ZZ$ coupling 
for highly detuned qubits~\cite{Xu2020a,Collodo2020a}. 
Thus, zero $ZZ$ coupling in the single-transmon coupler results in qubit-frequency crowding 
or crosstalk between qubits~\cite{flux-qubit-coupler}.

In this paper, we theoretically propose a new kind of tunable coupler, 
which we call a double-transmon coupler. 
Our coupler consists of two fixed-frequency transmons coupled through a common loop 
with an additional Josephson junction. 
We can control the coupling between the two coupler transmons 
by controlling the magnetic flux in the loop, 
and consequently tune the coupling strength between computational qubits. 
A remarkable feature of this coupler is that the $ZZ$ coupling vanishes 
even for highly detuned computational qubits, 
unlike the single-transmon coupler. 
Our numerical simulations indicate that this coupler allows us to achieve 
not only high two-qubit gate fidelities of over 99.99\% 
with a short gate time of 24~ns, but also no residual $ZZ$ coupling 
during idle time for highly detuned fixed-frequency transmons with detuning of 0.7~GHz. 
Thus, the double-transmon coupler is expected to be promising for improving
the performance of superconducting quantum computers.

\section{Design and mechanism}

\begin{figure}
\centering
	\includegraphics[width=8.5cm]{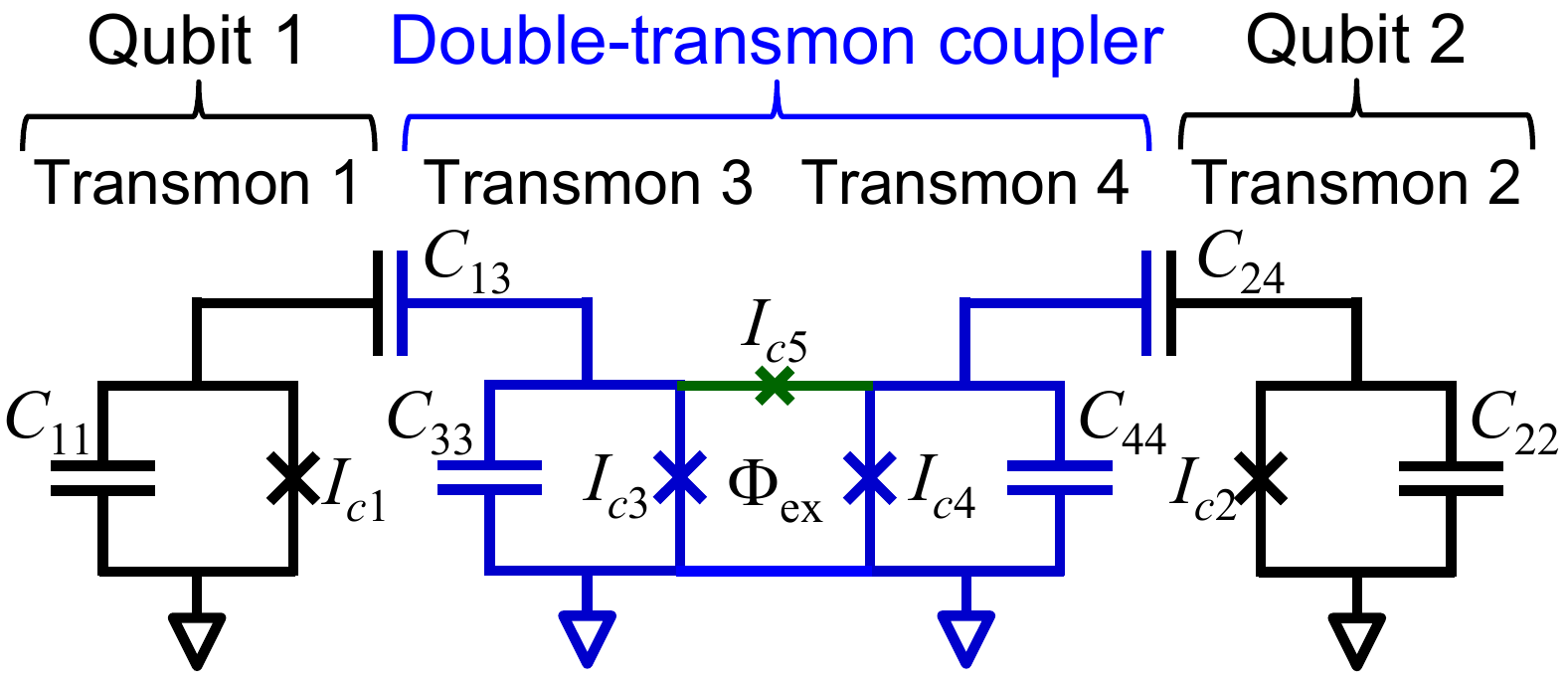}
	\caption{Diagram of the proposed double-transmon coupler 
	with two computational qubits. 
	Parasitic capacitances are considered in this work, but not shown for simplicity.}
	\label{fig-design}
\end{figure}

Figure~\ref{fig-design} 
shows a diagram of the proposed coupler. 
This consists of two fixed-frequency transmons [Transmons 3 and 4 in Fig.~\ref{fig-design}] 
coupled through a common loop with an additional Josephson junction, 
the critical current of which is smaller than that of the transmons. 
Two computational qubits [Transmons 1 and 2 in Fig.~\ref{fig-design}] 
are capacitively coupled to the coupler, as shown in Fig.~\ref{fig-design}.

The mechanism of this coupler is qualitatively explained 
from a classical point of view under rough approximations as follows~\cite{mechanism}. 
The Lagrangian describing the total system is given by ${L=K-V}$ with 
\begin{align}
K &= \sum_{i=1}^4 \frac{C_{ii}}{2} \dot{\phi}_i^2 
+ \sum_{i=1}^2 \sum_{j=i+1}^4 \frac{C_{ij}}{2} (\dot{\phi}_i - \dot{\phi}_j )^2 
+ \frac{C_{34}}{2} \dot{\phi}_5^2,
\label{eq-K}
\\
V &= -\sum_{i=1}^5 \hbar \omega_{Ji} \cos \varphi_i,
\label{eq-V}
\end{align}
where $\varphi_i$, ${\phi_i=\phi_0 \varphi_i}$, and ${\hbar \omega_{Ji}=\phi_0 I_{ci}}$ are, respectively, 
the phase difference, flux variable, and Josephson energy for the $i$th Josephson junction with critical current of $I_{ci}$~\cite{hbar}, 
the dots denote time derivatives, and we have included 
unwanted parasitic capacitances.

Neglecting the parasitic capacitances and 
using the constraint that ${\phi_5 = \phi_4 - \phi_3 - \Phi_{\mathrm{ex}}}$ 
($\Phi_{\mathrm{ex}}$ is the external flux in the loop)~\cite{constraint}, 
we can approximate $K$ and $V$ as
\begin{align}
K &\approx \sum_{i=1}^4 \frac{C_{ii}}{2} \dot{\phi}_i^2 
+ \frac{C_{13}}{2} (\dot{\phi}_1 - \dot{\phi}_3 )^2 
+ \frac{C_{24}}{2} (\dot{\phi}_2 - \dot{\phi}_4 )^2, 
\label{eq-Kapp}
\\
V &\approx -\sum_{i=1}^4 \hbar \omega_{Ji} \cos \varphi_i 
-\hbar \omega_{J5} \cos ( \varphi_4 - \varphi_3 - \Theta_{\mathrm{ex}} ),
\label{eq-Vapp}
\end{align}
where ${\Theta_{\mathrm{ex}}=\Phi_{\mathrm{ex}}/\phi_0}$.
Note that the two qubits are coupled only through 
the coupling between the two coupler transmons given by the last term in Eq.~(\ref{eq-Vapp}). 
This coupling can approximately be turned off 
by tuning the external flux $\Phi_{\mathrm{ex}}$, as follows.

Here we focus on the potential for the coupler $V_c$. 
In the transmon regime where Josephson energies are much larger 
than charging energies~\cite{Koch2007a}, 
low-energy states concentrate around a potential minimum, and hence 
the following second-order approximation is valid:
\begin{widetext}
\begin{align}
V_c
&=- \hbar \omega_{J3} \cos \varphi_3  - \hbar \omega_{J4} \cos \varphi_4  
- \hbar \omega_{J5} \cos (\varphi_4 - \varphi_3 - \Theta_{\mathrm{ex}} )
\nonumber
\\
&\approx
\frac{\hbar}{2} 
\begin{pmatrix}
\delta_3 & \delta_4
\end{pmatrix}
\begin{pmatrix}
\omega_{J3} \cos \varphi_3^{(0)} + 
\omega_{J5} \cos (\varphi_4^{(0)} -\varphi_3^{(0)} -\Theta_{\mathrm{ex}}) & 
-\omega_{J5} \cos (\varphi_4^{(0)} -\varphi_3^{(0)} -\Theta_{\mathrm{ex}}) 
\\
-\omega_5 \cos (\varphi_4^{(0)} -\varphi_3^{(0)} -\Theta_{\mathrm{ex}}) & 
\omega_{J4} \cos \varphi_4^{(0)} + 
\omega_{J5} \cos (\varphi_4^{(0)} -\varphi_3^{(0)} -\Theta_{\mathrm{ex}})
\end{pmatrix}
\begin{pmatrix}
\delta_3 \\ \delta_4
\end{pmatrix},
\label{eq-Vc}
\end{align}
\end{widetext}
where 
$\varphi_3^{(0)}$ and $\varphi_4^{(0)}$ are the phase differences minimizing $V_c$, 
${\delta_i = \varphi_i - \varphi_i^{(0)}}$, 
and constants have been dropped.
Note that when $\Theta_{\mathrm{ex}}$ is equal to $\Theta_{\mathrm{ex}}^{(0)}$ 
satisfying ${\cos (\varphi_4^{(0)} -\varphi_3^{(0)} -\Theta_{\mathrm{ex}}^{(0)})=0}$, 
nondiagonal elements of the matrix in Eq.~(\ref{eq-Vc}) vanish, and consequently 
the coupling between the two coupler transmons is turned off, as desired. 
This mechanism is substantially different from 
that of the single-transmon coupler~\cite{flux-qubit-coupler}.

\section{$ZZ$ coupling}

In order to accurately evaluate 
the properties of the double-transmon coupler, 
here we numerically investigate it using a fully quantum-mechanical model 
with finite parasitic capacitances. 
In this work, the qubits are assumed to be detuned. 
The qubit states are then well-defined by the energy eigenstates of the total Hamiltonian. 
However, there can be an unwanted correlated energy shift due to residual $ZZ$ coupling. 
The $ZZ$ coupling strength $\zeta_{ZZ}$ is defined as 
\begin{align}
\zeta_{ZZ} = \omega_{11} - (\omega_{10} + \omega_{01}),
\label{eq-ZZ}
\end{align}
where ${\omega_{ij}=E_{ij}/\hbar}$ is the frequency 
corresponding to the energy, $E_{ij}$, of the two-qubit state $|ij\rangle$. 
We also set the origin of energy as ${\omega_{00}=0}$.
When ${\zeta_{ZZ}=0}$, the two qubits are completely independent.

\begin{figure}
\centering
	\includegraphics[width=8.5cm]{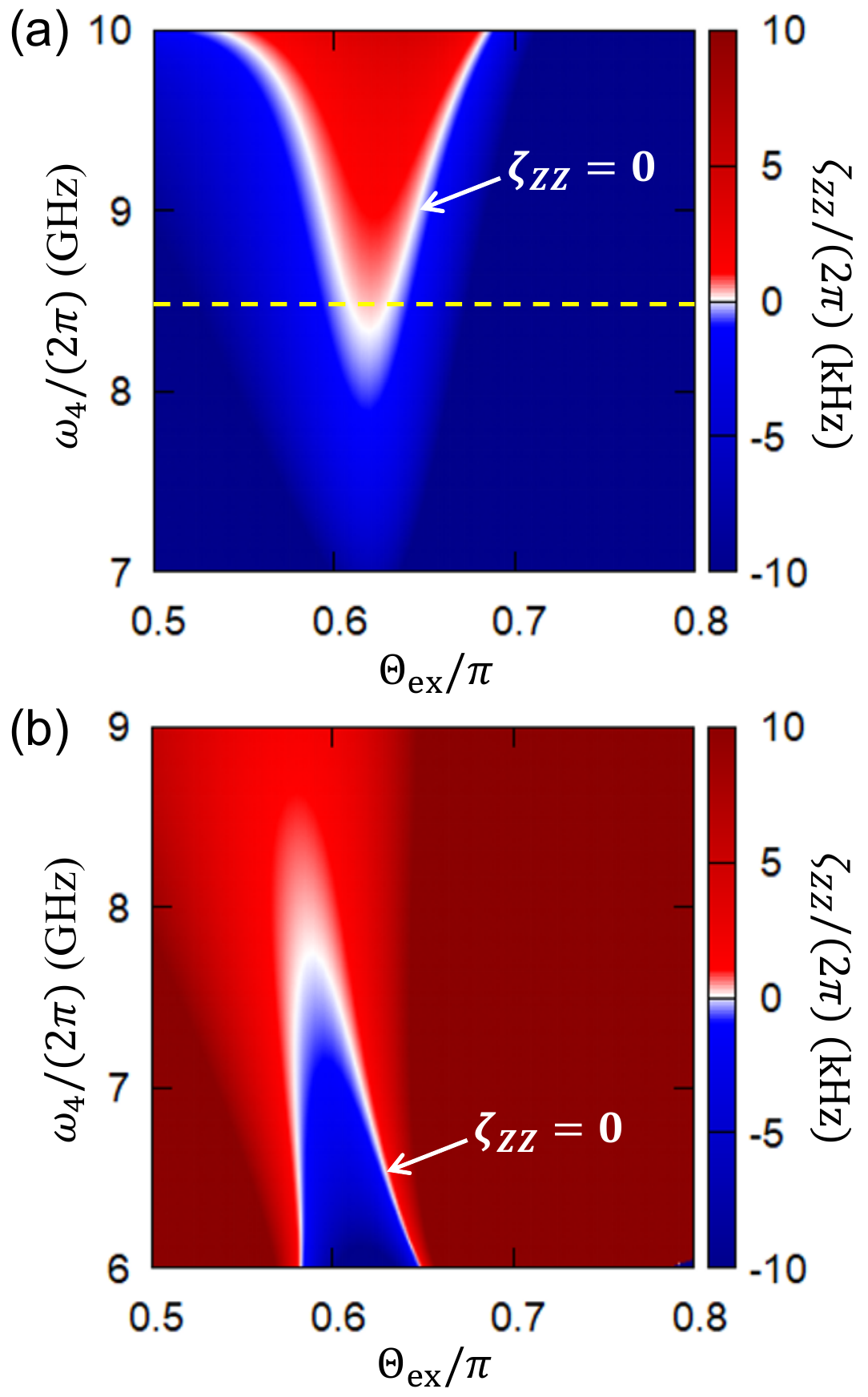}
	\caption{$ZZ$-coupling strength $\zeta_{ZZ}$ in the double-transmon coupler 
	as a function of $\Theta_{\mathrm{ex}}$ and 
	the design value of the Transmon-4 frequency denoted by $\omega_4$. 
	Design values of the other transmon frequencies are set as 
	${\omega_1/(2\pi)=5}$~GHz, ${\omega_2/(2\pi)=5.7}$~GHz (a) or 5.1~GHz (b), 
	and ${\omega_3/(2\pi)=7.2}$~GHz. 
	Detuning between the qubits is thus about 0.7~GHz in (a) and 0.1~GHz in (b). 
	Capacitances are set as ${C_{11}=C_{22}=C_{33}=C_{44}=60}$~fF, 
	${C_{13}=C_{24}=6}$~fF, ${C_{34}=1}$~fF, ${C_{14}=C_{23}=2C_{12}=0.05}$~fF. 
	Resultant anharmoniciteis of the four transmons are about 0.3~GHz (see Appendix~\ref{appendix-parameter}). 
	Thus, (a) is out of the straddling regime and (b) is in the straddling regime.
	Horizontal dashed line in (a) [${\omega_4/(2\pi)=8.5}$~GHz] indicates the situation 
	for two-qubit gate simulations presented in Fig.~\ref{fig-gate}.}
	\label{fig-ZZ}
\end{figure}

By numerically diagonalizing the quantum-mechanical Hamiltonian 
derived from the Lagrangian given by Eqs.~(\ref{eq-K}) and (\ref{eq-V}) (see Appendix~\ref{appendix-model}),
we evaluate $\zeta_{ZZ}$
for two situations: larger and smaller detunings between the qubits 
than the anharmonicities (Kerr coefficients) of the qubits, 
which are, respectively, called ``out of the straddling regime'' 
and ``in the straddling regime.''
The results in the two situations are, respectively, 
shown in Figs.~\ref{fig-ZZ}(a) and \ref{fig-ZZ}(b), 
where the parameters are set to experimentally feasible values~\cite{Collodo2020a}. 
From these results, it turns out that 
the double-transmon coupler can have the vanishing points of the $ZZ$ coupling
in both the regimes.
This is a remarkable feature of the double-transmon coupler, 
because for the conventional single-transmon coupler, 
the $ZZ$-coupling vanishing points exist only 
in the straddling regime~\cite{Stehlik2021a} (see Appendix~\ref{appendix-coupler}). 
It is also interesting that the coupler-transmon frequency required 
for the zero $ZZ$ coupling is lower bounded out of the straddling regime, 
but upper bounded in the stradding regime~\cite{bound}.

\section{Two-qubit gate}

\begin{figure}
\centering
	\includegraphics[width=8.5cm]{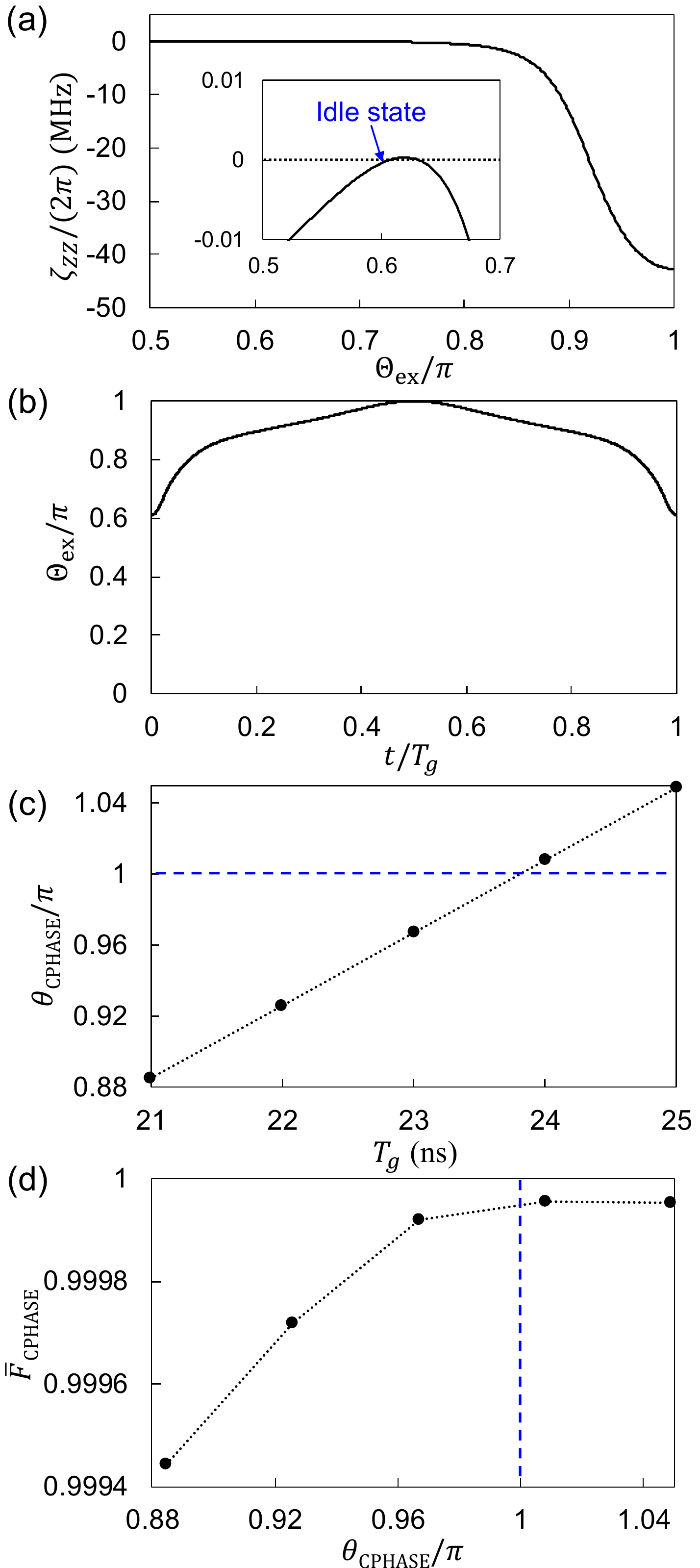}
	\caption{Two-qubit gate with the double-transmon coupler. 
	(a) $ZZ$-coupling strength $\zeta_{ZZ}$ on the dashed line in Fig.~\ref{fig-ZZ}(a) [${\omega_4/(2\pi)=8.5~\mathrm{GHz}}$]. 
	Inset is the magnification around the $ZZ$-coupling vanishing points. 
	(b) Flux pulse shape for the CPHASE gate ($T_g$: gate time). 
	(c) Rotation angle $\theta_{\mathrm{CPHASE}}$ of the CPHASE gate 
	with gate time of $T_g$.
	(d) Average CPHASE-gate fidelity $\bar{F}_{\mathrm{CPHASE}}$ corresponding to 
	$\theta_{\mathrm{CPHASE}}$. 
	In (c) and (d), dashed lines indicate ${\theta_{\mathrm{CPHASE}}=\pi}$, corresponding to the CZ gate.}
	\label{fig-gate}
\end{figure}

We evaluate two-qubit gate performance 
by numerical simulations with the parameter values in Fig.~\ref{fig-ZZ}(a) 
and ${\omega_4/(2\pi)=8.5}$~GHz [indicated by the horizontal dashed line 
in Fig.~\ref{fig-ZZ}(a)]~\cite{parameters}. 
The $\Theta_{\mathrm{ex}}$ dependence of $\zeta_{ZZ}$ for these parameter values is 
shown in Fig.~\ref{fig-gate}(a).
As shown in the inset, 
the $ZZ$ coupling vanishes at ${\Theta_{\mathrm{ex}} \simeq 0.61\pi}$ and $0.63\pi$. 
We thus define the qubit states by the energy eigenstates 
at ${\Theta_{\mathrm{ex}}=0.61\pi}$.
In other words, we set ${\Theta_{\mathrm{ex}}=0.61\pi}$ during idle time, 
as indicated in Fig.~\ref{fig-gate}(a).

In Fig.~\ref{fig-gate}(a), 
it is also notable that ${|\zeta_{ZZ}|/(2\pi)}$ becomes as large as 40~MHz 
at ${\Theta_{\mathrm{ex}}=\pi}$. 
This property can be used for a fast two-qubit gate called the controlled-phase (CPHASE) gate 
including the controlled-$Z$ (CZ) gate~\cite{Xu2020a,Stehlik2021a,Collodo2020a}, 
where $\zeta_{ZZ}$ is adiabatically increased and then decreased 
by controlling the external flux $\Phi_{\mathrm{ex}}$.
The flux pulse shape in the present simulations is shown in Fig.~\ref{fig-gate}(b), 
which is designed according to 
a technique for reducing nonadiabatic errors~\cite{Martinis2014a} (see Appendix~\ref{appendix-pulse}).
The simulation results are shown in Figs.~\ref{fig-gate}(c) and \ref{fig-gate}(d).

Figure~\ref{fig-gate}(c) 
shows that the rotation angle, $\theta_{\mathrm{CPHASE}}$, of the CPHASE gate 
increases linearly as the gate time $T_g$ increases. 
The CZ gate corresponding to ${\theta_{\mathrm{CPHASE}}=\pi}$ can be achieved 
when ${T_g \simeq 24}$~ns, as indicated by the horizontal dashed line in Fig.~\ref{fig-gate}(c). 
The average fidelity of the CPHASE gate is shown in Fig.~\ref{fig-gate}(d) (see Appendix~\ref{appendix-fidelity}), 
suggesting that the CZ-gate fidelity, 
indicated by the vertical dashed line in Fig.~\ref{fig-gate}(d), will surpass 99.99\%. 
Thus, the double-transmon coupler allows us to simultaneously achieve fast high-fidelity two-qubit gates 
and no residual coupling during idle time for highly detuned qubits~\cite{spectator}.
The infidelity is mainly due to leakage errors caused by nonadiabatic transitions 
from $|01\rangle$ and $|11\rangle$ to higher levels outside the qubit subspace (see Appendix~\ref{appendix-pulse}).
For instance, when the gate time is 24~ns, 
20\% and 73\% of the average infidelity are due to the leakage errors 
from ${|01\rangle}$ and ${|11\rangle}$, respectively.

\section{Flux noise}

Although the qubits are fixed-frequency transmons, 
the qubit frequencies vary a little depending on the flux in the coupler (see Appendix~\ref{appendix-model}). 
Here we examine the influence of the flux noise on the qubit coherence.

The coherence time $T_2$ for Qubit~1 in terms of the flux noise is formulated as~\cite{Koch2007a}
\begin{align}
T_2 \approx \left| A_{\Phi} \frac{\partial \omega_{10}}{\partial \Phi_{\mathrm{ex}}} \right|^{-1}
=
\left| 2\pi \frac{A_{\Phi}}{\Phi_0} \frac{\partial \omega_{10}}{\partial \Theta_{\mathrm{ex}}} \right|^{-1},
\label{eq-T2}
\end{align}
where the coefficient $A_{\Phi}$ is typically $10^{-5} \Phi_0$~\cite{hbar,Koch2007a} 
($T_2$ for Qubit~2 is given similarly). 
Using the parameter values in Fig.~\ref{fig-gate} and ${A_{\Phi}=10^{-5} \Phi_0}$, 
$T_2$ for Qubits~1 and 2 
are numerically estimated to be about 260~$\mu$s and 430~$\mu$s, respectively, in idle state.
These long coherence times suggest the robustness of the proposed scheme against flux noise.

During the two-qubit gate, $\Theta_{\mathrm{ex}}$ changes from $0.61\pi$ to $\pi$.
In this range of $\Theta_{\mathrm{ex}}$, 
the minimum values of $T_2$ estimated as above are 30~$\mu$s and 5$\mu$s, respectively, 
for Qubits~1 and 2.
Even in the worst-case scenario where the coherence time is assumed to be 5$\mu$s, 
the infidelity of the CPHASE gate with the gate time of 24~ns 
may increase to about 0.5\%, which is still small. 
This rough estimation suggests that the flux noise may not degrade the gate performance very much.

\section{Conclusions}

We have theoretically proposed a new kind of tunable coupler 
for superconducting quantum computers. 
We call this a double-transmon coupler, because this consists of two fixed-frequency transmons 
coupled through a common loop with an additional Josephson junction. 
We have numerically found that by tuning the external flux in the loop, 
residual $ZZ$ coupling vanishes even for highly detuned computational qubits, 
in contrast to the conventional single-transmon coupler. 
Numerical simulations have also shown that the proposed coupler enables two-qubit gates 
with high fidelity of over 99.99\% 
and a short gate time of 24~ns. 
The next step is experimental realization of this proposal, 
where relaxation and decoherence in transmons will degrade the performance. 
However, from its short gate time (24~ns) and 
recently reported long coherence times of transmons (over $300~\mu$s)~\cite{Place2021a,Wang2022a}, 
the proposed coupler is expected to achieve high two-qubit gate fidelity.
Another important issue in experiments is the unwanted deviation of critical currents of Josephson junctions
from design values.
The precision of the critical currents is known to be about 2\%, 
though this can be reduced by laser annealing~\cite{Hertzberg2021a}.
The effects of the critical-current deviation on the coupler performance are left as an important issue 
for future work.

\begin{appendix}

\section{Quantum-mechanical model}
\label{appendix-model}

Using the constraint that ${\phi_5 = \phi_4 - \phi_3 - \Phi_{\mathrm{ex}}}$, 
the kinetic energy term in Eq.~(1) can be expressed as 
\begin{align}
K = \frac{1}{2} \dot{\boldsymbol{\phi}}^T M \dot{\boldsymbol{\phi}}- \mathbf{q}^T \dot{\boldsymbol{\phi}},
\label{eq-K2}
\end{align}
where $\dot{\boldsymbol{\phi}}^T=(\dot{\phi}_1~\dot{\phi}_2~\dot{\phi}_3~\dot{\phi}_4)$, 
$\mathbf{q}^T=(0~0~-C_{34} \dot{\Phi}_{\mathrm{ex}}~C_{34} \dot{\Phi}_{\mathrm{ex}})$, and 
$M$ is a capacitor matrix.
The canonical conjugate variables for the flux variables, namely, charge variables $\mathbf{Q}$, 
and the Hamiltonian are obtained as
\begin{align}
\mathbf{Q} &= \frac{\partial L}{\partial \dot{\boldsymbol{\phi}}}=M \dot{\boldsymbol{\phi}} - \mathbf{q},
\label{eq-Q}
\\
H &= \mathbf{Q}^T \dot{\boldsymbol{\phi}} - L
= \frac{1}{2} \mathbf{Q}^T M^{-1} \mathbf{Q} + \mathbf{q}^T M^{-1} \mathbf{Q} + V.
\label{eq-H}
\end{align}
Introducing the Cooper-pair number variables as $\mathbf{n}=\mathbf{Q}/(2\mathrm{e})$ 
(e is the elementary charge), 
$H$ is rewritten as
\begin{align}
H=4\hbar \mathbf{n}^T W \mathbf{n} + \hbar \frac{\dot{\Theta}_{\mathrm{ex}}}{\omega_{\mathrm{C}34}} 
(0~0~-1~1) W \mathbf{n} + V,
\label{eq-H2}
\end{align}
where ${\displaystyle \hbar W=\frac{\mathrm{e}^2}{2} M^{-1}}$ and 
${\displaystyle \hbar \omega_{\mathrm{C}34}=\frac{\mathrm{e}^2}{2C_{34}}}$ 
have been introduced as frequency parameters.

The variables are quantized by the commutation relation 
${[\hat{\varphi}_i, \hat{n}_j ]=\mathrm{i} \delta_{ij}}$ as follows.
$\hat{n}_i$ is represented by ${-\mathrm{i} \frac{\partial}{\partial \varphi_i}}$ 
and the eigenfunction of $\hat{n}_i$ is proportional to 
$e^{\mathrm{i} n_i \varphi_i}$. 
In the basis of these eigenfunctions, 
we have the following matrix representation of operators:
\begin{align}
\hat{n}_i &= 
\begin{pmatrix}
-N & & \\
 & \ddots & \\
 & & N
\end{pmatrix},
\label{eq-n}
\\
\cos \hat{\varphi}_i &= 
\frac{1}{2}
\begin{pmatrix}
 & 1 & & \\
1 & & \ddots & \\
 & \ddots & & 1 \\
 & & 1 &
\end{pmatrix},
\label{eq-cos}
\\
\sin \hat{\varphi_i} &= 
\frac{1}{2\mathrm{i}} 
\begin{pmatrix}
 & -1 & & \\
1 & & \ddots & \\
 & \ddots & & -1 \\
 & & 1 & 
\end{pmatrix},
\label{eq-sin}
\end{align}
where we have truncated the number of Cooper pairs at $\pm N$.

Since the total system is composed of four subsystems (transmons), 
each operator in Eq.~(\ref{eq-H2}) is represented 
by a tensor product of four operators, 
such as ${\hat{n}_1 \otimes \hat{I}_2 \otimes \hat{I}_3 \otimes \hat{I}_4}$, 
where $\hat{I}_i$ is the identity operator for the $i$th subsystem of 
$\hat{\varphi}_i$ and $\hat{n}_i$. 
From the addition theorem, 
${\cos (\varphi_4 - \varphi_3 - \Theta_{\mathrm{ex}} )}$ can be expressed 
as $\cos \Theta_{\mathrm{ex}} 
\left[ \hat{I}_1 \otimes \hat{I}_2 \otimes 
(\cos \hat{\varphi}_3 \otimes \cos \hat{\varphi}_4 
+ \sin \hat{\varphi}_3 \otimes \sin \hat{\varphi}_4 ) \right]
+ \sin \Theta_{\mathrm{ex}} \left[ \hat{I}_1 \otimes \hat{I}_2 \otimes 
(\cos \hat{\varphi}_3 \otimes \sin \hat{\varphi}_4 
- \sin \hat{\varphi}_3 \otimes \cos \hat{\varphi}_4 ) \right]$.

In the matrix representation, $\hat{I}_i$ is given by 
the ${(2N+1)\times (2N+1)}$ unit matrix and the tensor product $\otimes$ is replaced 
by the Kronecker product of matrices. 
Thus, we obtain a ${(2N+1)^4 \times (2N+1)^4}$ matrix representation of the Hamiltonian 
in Eq.~(\ref{eq-H2}). 
In this work, we choose ${N=10}$ for sufficient convergence of energies.

\begin{figure}
\centering
	\includegraphics[width=8.5cm]{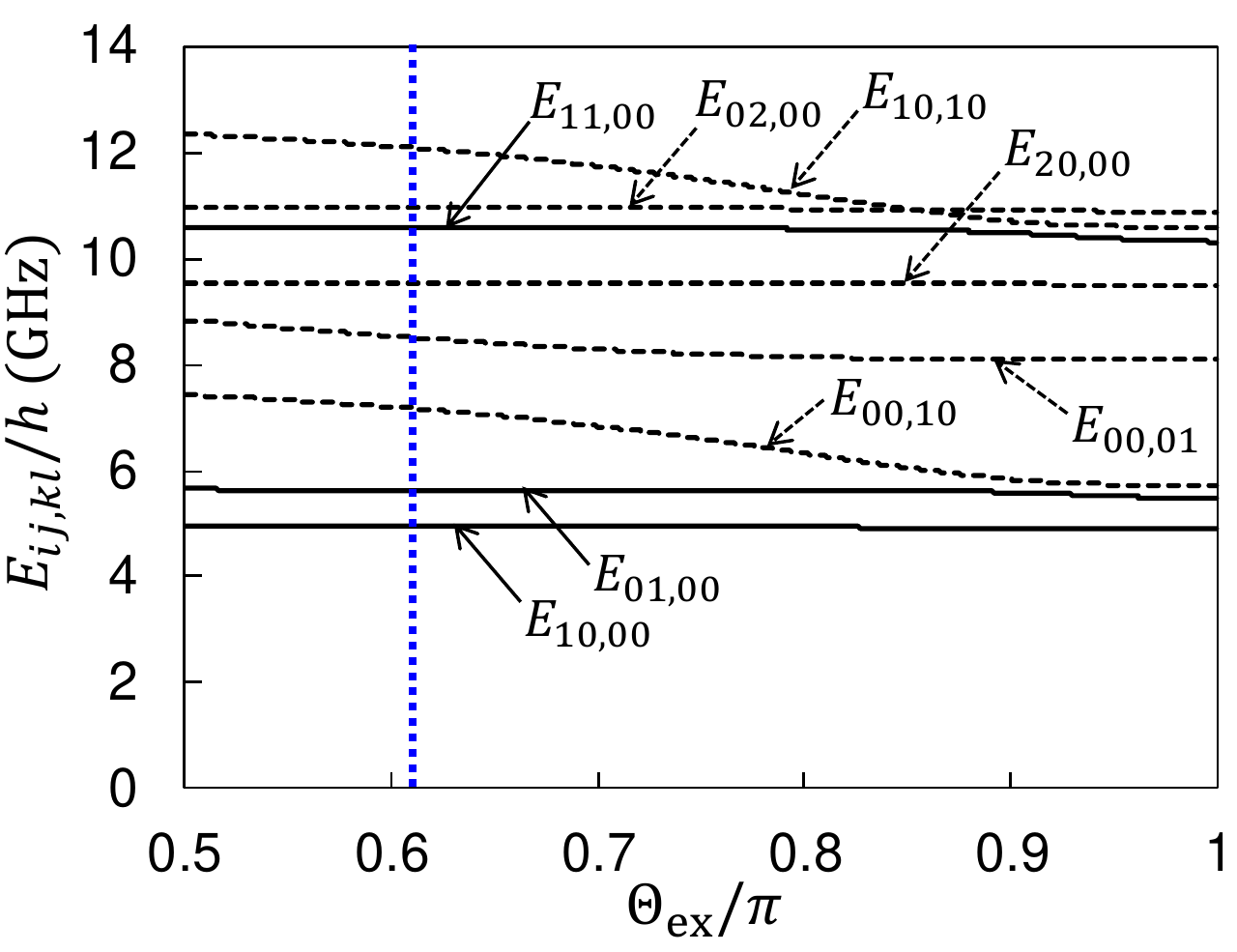}
	\caption{Energies, $E_{ij,kl}$, of $|ij\rangle |kl\rangle$ in the case of Fig.~3. 
	($h$ is the Planck constant.) 
	Solid curves correspond to two-qubit states: 
	${\omega_{ij}=E_{ij,00}/\hbar}$ (${\omega_{00}=E_{00,00}=0}$).
	Vertical dotted line indicates the idle state (${\Theta_{\mathrm{ex}}=0.61\pi}$).}
	\label{fig-energy}
\end{figure}

Numerically diagonalizing the Hamiltonian matrix with ${\dot{\Theta}_{\mathrm{ex}}=0}$, 
we can obtain the energies, $E_{ij,kl}$, of the state $|ij\rangle |kl\rangle$, 
where $|ij\rangle$ and $|kl\rangle$ denote the qubit state (Qubits~1 and 2) and 
the coupler state (Transmons~3 and 4), respectively.
These energies lead to the $ZZ$-coupling strength $\zeta_{ZZ}$ in Figs.~2 and 3(a). 
For example, the energies in the case of Fig.~3 are shown in Fig.~\ref{fig-energy}.
The qubit energies vary a little depending on $\Theta_{\mathrm{ex}}$. 
The slopes of the qubit-energy curves give the estimated values of $T_2$. 
Figure~\ref{fig-energy} also shows that the levels corresponding to the coupler excited states 
largely change by the flux and the other levels do not, as expected.

The simulation results of the CPHASE gate in Fig.~3 are obtained 
by numerically solving the Schr\"{o}dinger equation with the Hamiltonian matrix. 
The average fidelity $\bar{F}_{\mathrm{CPHASE}}$ and 
rotation angle $\theta_{\mathrm{CPHASE}}$ of the CPHASE gate 
in Fig.~3 are obtained as explained in the next section.

\section{Average fidelity and rotation angle of the CPHASE gate}
\label{appendix-fidelity}

The average gate fidelity is a standard metric for evaluating the performance of quantum gates. 
This is defined by averaging gate fidelities over uniformly distributed initial states. 
The average fidelities in Fig.~3(d) are obtained using the formula in Ref.~\citenum{Kueng2016a}, 
which is an extension of the formula in Ref.~\citenum{Nielsen2002a} 
to cases where there exist leakage errors and 
the norm of the qubit-subspace vector is not preserved. 
In the case of two-qubit gates, 
the formula for the average fidelity $\bar{F}$ is given by
\begin{align}
\bar{F} = \frac{\left| \mathrm{tr} \left( U_{\mathrm{id}}^{\dagger} U' \right) \right|^2
+ \mathrm{tr} \left( U^{\prime \dagger} U' \right)}{20},
\label{eq-F}
\end{align}
where $U_{\mathrm{id}}$ is a ${4\times 4}$ unitary matrix corresponding to the ideal gate operation 
and $U'$ is a ${4\times 4}$ matrix defined as follows. 
Suppose that we simulate the gate operation on four initial states 
each of which is one of the four two-qubit basis vectors denoted by 
$|\psi_{ij} \rangle$ (${i, j = 0, 1}$). 
Using the resultant vectors $|\psi'_{ij} \rangle$, 
$U'$ is defined as 
${U'_{2i+j,2i'+j'}=\langle \psi_{ij} | \psi'_{i'j'} \rangle}$. 
Note that $U'$ is not a unitary matrix in general because of leakage errors. 
In the case of the CPHASE gate, 
we define $U_{\mathrm{id}}$ as 
${U_{\mathrm{id}} = 
\mathrm{diag}(e^{\mathrm{i} \theta_0}, e^{\mathrm{i} \theta_1}, 
e^{\mathrm{i} \theta_2}, e^{\mathrm{i} \theta_3})}$, 
where $\mathrm{diag}(\cdots)$ represents a diagonal matrix and 
${e^{\mathrm{i} \theta_k}=U'_{k,k}/|U'_{k,k}|}$. 
By eliminating the overall phase factor and single-qubit phase rotations from $U_{\mathrm{id}}$, 
we define the rotation angle of the CPHASE gate as 
${\theta_{\mathrm{CPHASE}} = \theta_3 - \theta_1 - \theta_2 + \theta_0}$.

\section{Flux pulse shape}
\label{appendix-pulse}

As shown in Fig.~\ref{fig-energy},
higher energy levels $E_{00,10}$ and $E_{10,10}$ 
approach the qubit levels $E_{01,00}$ and $E_{11,00}$, respectively, 
around ${\Theta_{\mathrm{ex}}=\pi}$.
Thus, the infidelity of the CPHASE gate is mainly due to 
the leakage errors from $|01\rangle |00\rangle$ to $|00\rangle |10\rangle$
and from $|11\rangle |00\rangle$ to $|10\rangle |10\rangle$.
To reduce these leakage errors, 
we design the flux pulse shape based on the technique proposed in Ref.~\citenum{Martinis2014a}. 
Here we explain how we designed the pulse shape shown in Fig.~\ref{fig-gate}(b).

\begin{figure}
\centering
	\includegraphics[width=8.5cm]{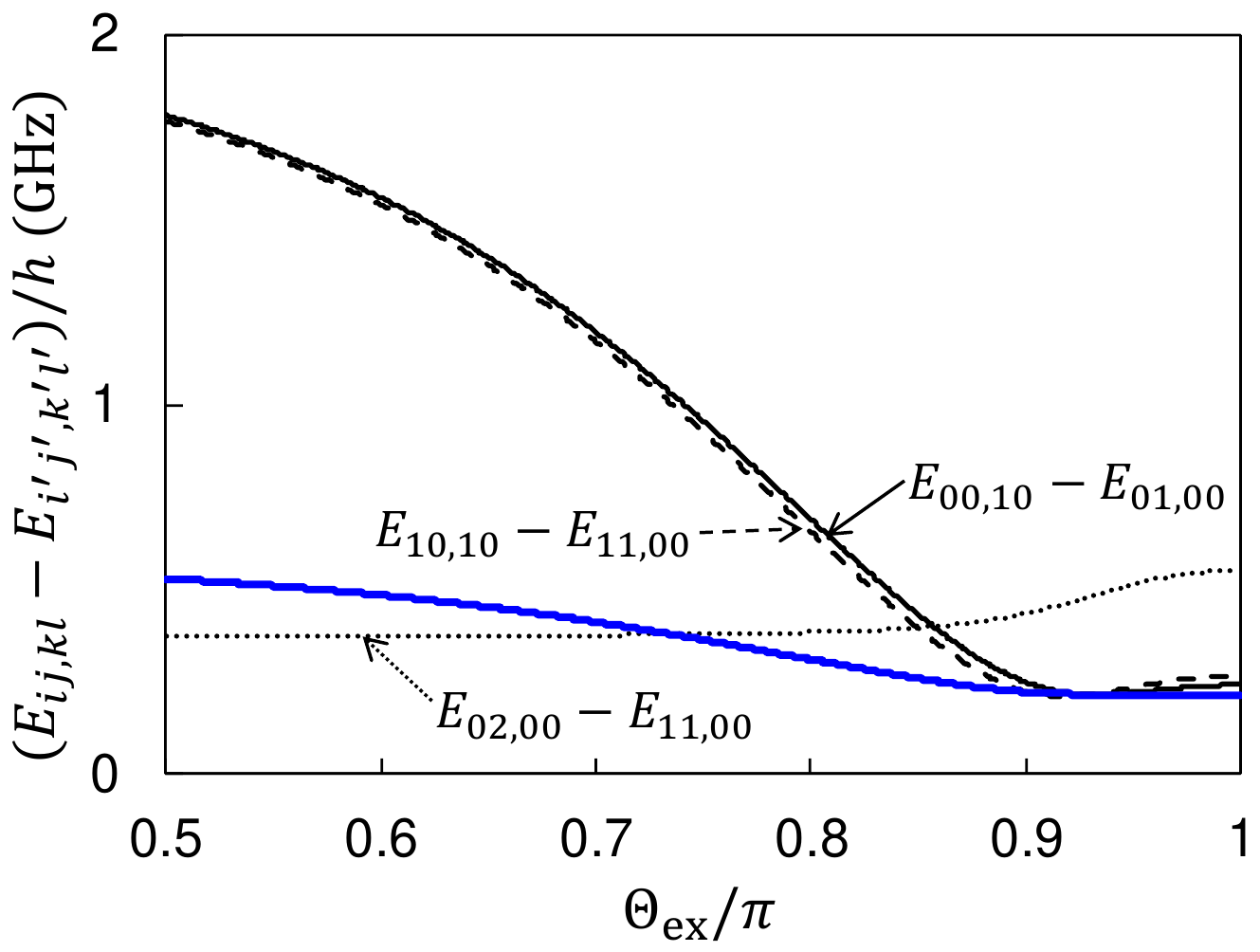}
	\caption{Energy gaps. 
	Thin solid curve (energy gap between $|01\rangle |00\rangle$ and $|00\rangle |10\rangle$) is used as $\omega_{\mathrm{gap}}$
	required for designing the flux pulse shape for the CPHASE gate.
	Bold solid curve (in blue), which is given by Eq.~(\ref{eq-gap}), 
	is used as $\omega_{\mathrm{gap}}$ in ${t(s)=\! \int_0^s \omega_{\mathrm{gap}}(s')^{-1} ds'}$.}
	\label{fig-gap}
\end{figure}

The technique is based on the two-level system, $|g\rangle$ and $|e\rangle$, 
with a constant coupling rate $g$ and a time-dependent detuning $\Delta (t)$, 
where the energy gap between the two energy eigenstates of this sysmte is 
given by ${\hbar \omega_{\mathrm{gap}}=\hbar \sqrt{\Delta^2+4g^2}}$.
We first focus on the two levels of $|01\rangle |00\rangle$ and $|00\rangle |10\rangle$.
In this case, 
we have ${\omega_{\mathrm{gap}}=(E_{00,10}-E_{01,00})/\hbar}$, and $2g$ is given by the minimum of $\omega_{\mathrm{gap}}$.
The energy gap is shown in Fig.~\ref{fig-gap} together with other energy gaps.
From this, we obtain $g$ and the $\Theta_{\mathrm{ex}}$ dependences of $\omega_{\mathrm{gap}}$ and $\Delta$.

The two energy eigenstates are expressed as 
${\cos (\theta /2)|g\rangle - \sin (\theta /2)|e\rangle}$ 
and ${\sin (\theta /2)|g\rangle + \cos (\theta /2)|e\rangle}$ 
with ${\theta = \arctan (2g/\Delta)}$. 
Then, the nonadiabatic error probability $P_e$ is approximately formulated as~\cite{Martinis2014a}
\begin{align}
P_e &\approx \frac{1}{4} \! \left|
\int_0^{T_g} \! \frac{d\theta}{dt} e^{-i \! \int_0^t \omega_{\mathrm{gap}}(t') dt'} dt
\right|^2
= \frac{1}{4} \! \left|
\int_0^{s_f} \! \frac{d\theta}{ds} e^{-is} ds
\right|^2,
\label{eq-Pe}
\end{align}
where we have introduced a dimensionless time defined as
${s(t)=\! \int_0^t \omega_{\mathrm{gap}}(t') dt'}$
[$s_f$ is defined as ${s_f=s(T_g)}$].
In this work, we set $\theta (s)$ as~\cite{Martinis2014a} 
\begin{align}
\theta (s) &= \theta_0 + (\theta_1 - \theta_0) 
\nonumber \\
&\times 
\frac{\displaystyle \left[ \cos \! \left( 2\pi \frac{s}{s_f} \right) \! -1 \right] \! 
+ \frac{A}{2} \! \left[ \cos \! \left( 4\pi \frac{s}{s_f} \right) \! -1 \right]}{-2},
\label{eq-theta}
\end{align}
where ${\theta_0 = \theta (0)= \theta (s_f)}$ and ${\theta_1 = \theta (s_f/2)}$ are 
$\theta$ corresponding to 
$\Theta_{\mathrm{ex}}=0.61\pi$ and $\pi$, respectively, 
and the coefficient $A$ is set as ${A=-0.17}$ to achieve small $P_e$ and short $T_g$~\cite{comment-A}.
Using this $\theta (s)$, ${t(s)=\! \int_0^s \omega_{\mathrm{gap}}(s')^{-1} ds'}$, and
the $\Theta_{\mathrm{ex}}$ dependences of $\omega_{\mathrm{gap}}$ and $\Delta$, 
we obtain the corresponding pulse shape $\Theta_{\mathrm{ex}} (t)$.

However, we found that this pulse shape leads to relatively high leakage error probabilities from $|11\rangle |00\rangle$,
though the energy gap between $|11\rangle |00\rangle$ and $|10\rangle |10\rangle$ 
is close to that between $|01\rangle |00\rangle$ and $|00\rangle |10\rangle$, as shown in Fig.~\ref{fig-gap}.
The leakage errors may be due to the smaller energy gap between $|11\rangle |00\rangle$ and $|02\rangle |00\rangle$ 
around $\Theta_{\mathrm{ex}}=0.61\pi$. 
Also, slower change of $\Theta_{\mathrm{ex}}$ around $\Theta_{\mathrm{ex}}=\pi$ may be more desirable 
for increasing the rotation angle, because the $ZZ$-coupling strength becomes maximum there.
Inspired by these, 
we modified the energy gap used for ${t(s)=\! \int_0^s \omega_{\mathrm{gap}}(s')^{-1} ds'}$ as 
\begin{align}
\left\{
\begin{matrix}
0.2 \left[ \left( E_{00,10}-E_{01,00} \right)/\hbar -2g \right] +2g & \cdots & \Theta_{\mathrm{ex}} \le \Theta_{\mathrm{ex}}^{(g)},
\\
2g & \cdots & \Theta_{\mathrm{ex}} > \Theta_{\mathrm{ex}}^{(g)},
\end{matrix}
\right.
\label{eq-gap}
\end{align}
which is shown by the bold solid curve (in blue) in Fig.~\ref{fig-gap}
[$\Theta_{\mathrm{ex}}^{(g)}$ is $\Theta_{\mathrm{ex}}$ 
satsfying $(E_{00,10}-E_{01,00})/\hbar =2g$].
Thus, we obtain the pulse shape shown in Fig.~\ref{fig-gate}(b).

\section{Parameter values in numerical studies}
\label{appendix-parameter}

The parameter values used for the present numerical studies are set as follows. 
Note that the present system can be regarded as a network of capacitively coupled four transmons, 
except for the interaction between Transmons~3 and 4 
through the additional Josephson junction given by the last term in Eq.~(4). 
The transmon network is quantized by the standard method using bosonic operators~\cite{Blais2021a}. 
Its Hamiltonian is given by
\begin{align}
\hat{H}' =& 
\sum_{i=1}^4 
\left( \hbar \omega_i \hat{a}_i^{\dagger} \hat{a}_i 
- \hbar \frac{W_{ii}}{2} \hat{a}_i^{\dagger 2} \hat{a}_i^2 \right) 
\nonumber \\
&+ \sum_{i=1}^3 \sum_{j=i+1}^4 \hbar g_{ij} 
\left( \hat{a}_i^{\dagger} \hat{a}_j 
+ \hat{a}_j^{\dagger} \hat{a}_i \right),
\label{eq-Hprime}
\\
\omega_i  =& \sqrt{8 W_{ii} \omega_{Ji}} - W_{ii},
\label{eq-w}
\\
g_{ij} =& \frac{W_{ij}}{2} \sqrt{\frac{(\omega_i + W_{ii})(\omega_j + W_{jj} )}{W_{ii} W_{jj}}},
\label{eq-g}
\end{align}
where $\hat{a}_i$ and $\hat{a}_i^{\dagger}$ are 
the annihilation and creation operators, respectively, for the $i$th transmon. 
In this work, we set the transmon frequencies $\omega_i$ and capacitances as design values, 
as given in Fig.~2. 
The parasitic capacitances not shown in Fig.~\ref{fig-design} are ideally set to zero, but 
this is impossible in actual experiments. 
Among the parasitic capacitances, $C_{34}$ is set to a relatively large value, because 
Transmons~3 and 4 are directly coupled via the Josephson junction with critical current of $I_{c5}$. 
On the other hand, the other parasitic capacitances, 
which comprise nonadjacent transmons, are set to small values, 
which may be feasible by placing the nonadjacent transmons as far from each other as possible.
The other parameters are determined by their definitions, 
together with $\omega_{J5} = (\omega_{J3} + \omega_{J4})/8$ 
(a quarter of the mean value of $\omega_{J3}$ and $\omega_{J4}$). 
Table~\ref{tableS1} summarizes the design values (shown in bold) and 
resultant other parameter values used in this work. 
The anharmonicities (Kerr coefficients) of the transmons are given by $W_{ii}$, 
which are about 0.3~GHz.

\begin{table}[h]
	\caption{Parameter setting for the present numerical studies.
	Bold values are design values. 
	The others are calculated from their definitions using the design values. }
	
	\begin{minipage}{4cm}
	\begin{tabular*}{4cm}{wl{2.7cm}wr{1cm}}
	\hline
	$\omega_1/(2\pi)$~(GHz) & \textbf{5} \\
	$\omega_2/(2\pi)$~(GHz) & \textbf{5.7} \\
	$\omega_3/(2\pi)$~(GHz) & \textbf{7.2} \\
	$\omega_4/(2\pi)$~(GHz) & \textbf{8.5} \\
	\hline
	\\
	\hline
	$C_{11}$~(fF) & \textbf{60} \\
	$C_{12}$~(fF) & \textbf{0.025} \\
	$C_{13}$~(fF) & \textbf{6} \\
	$C_{14}$~(fF) & \textbf{0.05} \\
	$C_{22}$~(fF) & \textbf{60} \\
	$C_{23}$~(fF) & \textbf{0.05} \\
	$C_{24}$~(fF) & \textbf{6} \\
	$C_{33}$~(fF) & \textbf{60} \\
	$C_{34}$~(fF) & \textbf{1} \\
	$C_{44}$~(fF) & \textbf{60} \\
	\hline
	\\
	\hline
	$g_{12}/(2\pi)$~(MHz) & 1.7 \\
	$g_{13}/(2\pi)$~(MHz) & 239 \\
	$g_{14}/(2\pi)$~(MHz) & 5.7 \\
	$g_{23}/(2\pi)$~(MHz) & 6.5 \\
	$g_{24}/(2\pi)$~(MHz) & 270 \\
	$g_{34}/(2\pi)$~(MHz) & 57 \\
	\hline
	\end{tabular*}
	\end{minipage}
	\hfill
	\begin{minipage}{4cm}
	\begin{tabular*}{4cm}{wl{2.7cm}wr{1cm}}
	\hline
	$W_{11}/(2\pi)$~(MHz) & 296 \\
	$W_{12}/(2\pi)$~(MHz) & 0.19 \\
	$W_{13}/(2\pi)$~(MHz) & 26.5 \\
	$W_{14}/(2\pi)$~(MHz) & 0.63 \\
	$W_{22}/(2\pi)$~(MHz) & 296 \\
	$W_{23}/(2\pi)$~(MHz) & 0.63 \\
	$W_{24}/(2\pi)$~(MHz) & 26.5 \\
	$W_{33}/(2\pi)$~(MHz) & 291 \\
	$W_{34}/(2\pi)$~(MHz) & 4.42 \\
	$W_{44}/(2\pi)$~(MHz) & 291 \\
	\hline
	\\
	\hline
	$\omega_{J1}/(2\pi)$~(GHz) & 11.9 \\
	$\omega_{J2}/(2\pi)$~(GHz) & 15.2 \\
	$\omega_{J3}/(2\pi)$~(GHz) & 24.1 \\
	$\omega_{J4}/(2\pi)$~(GHz) & 33.2 \\
	$\omega_{J5}/(2\pi)$~(GHz) & 7.2 \\
	\hline
	\\
	\hline
	$I_{c1}/(2\pi)$~(nA) & 23.9 \\
	$I_{c2}/(2\pi)$~(nA) & 30.6 \\
	$I_{c3}/(2\pi)$~(nA) & 48.5 \\
	$I_{c4}/(2\pi)$~(nA) & 66.8 \\
	$I_{c5}/(2\pi)$~(nA) & 14.4 \\
	\hline
	\end{tabular*}
	\end{minipage}
	\label{tableS1}
\end{table}

\section{Single-transmon coupler}
\label{appendix-coupler}

\begin{figure}[h]
\centering
	\includegraphics[width=7.5cm]{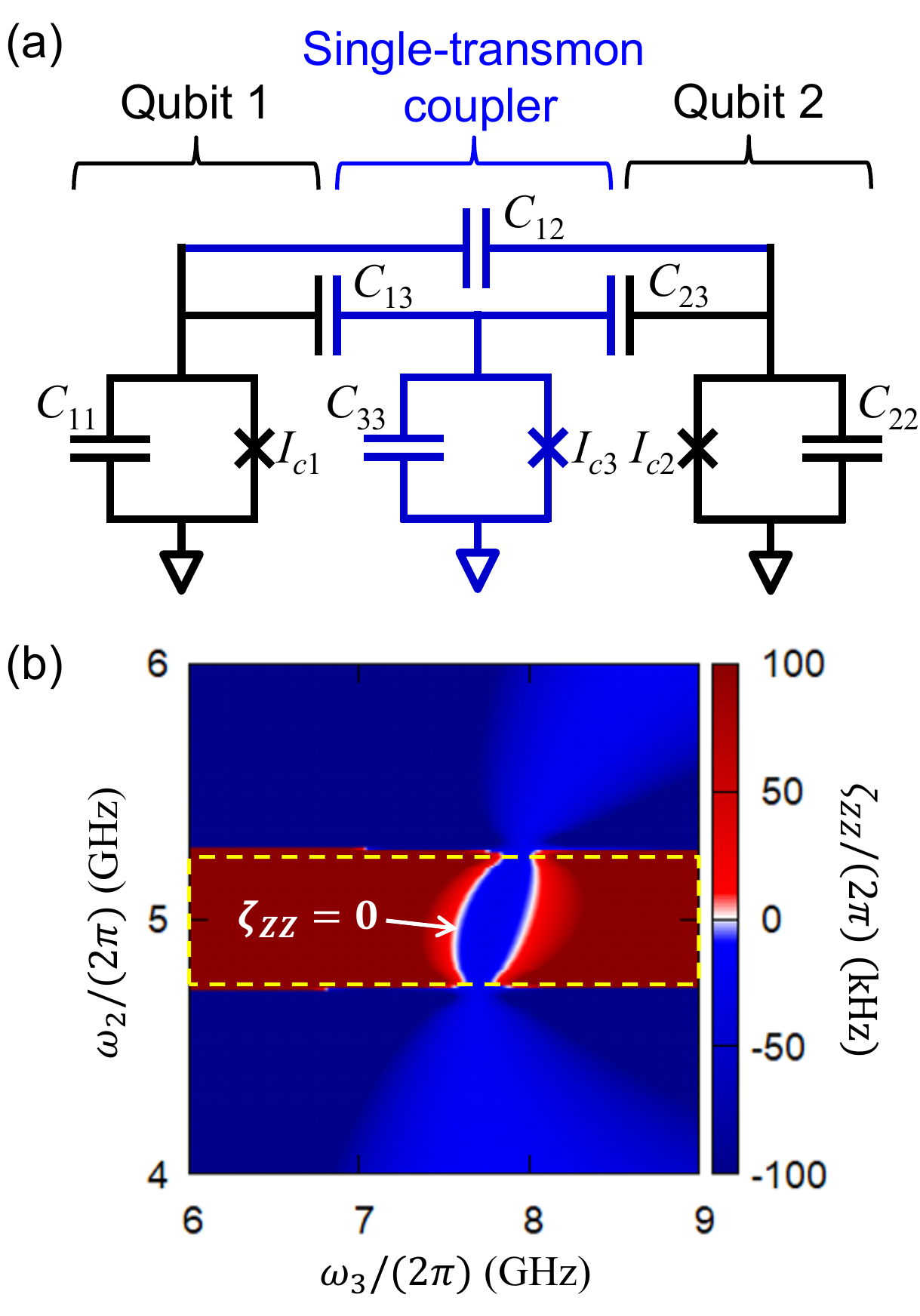}
	\caption{Single-transmon coupler. 
	(a) Simplified diagram. 
	(b) $ZZ$-coupling strength $\zeta_{ZZ}$ with 
	${\omega_1/(2\pi)=5~\mathrm{GHz}}$, 
	$W_{11}/(2\pi)=W_{22}/(2\pi)=W_{33}/(2\pi)=250~\mathrm{MHz}$, 
	$g_{13}/(2\pi)=g_{23}/(2\pi)=250~\mathrm{MHz}$, 
	$g_{12}/(2\pi)=25~\mathrm{MHz}$. 
	The dashed yellow box indicates the region in the straddling regime.}
	\label{fig-coupler}
\end{figure}

The conventional single-transmon coupler is shown in Fig.~\ref{fig-coupler}(a), 
in which the dc SQUID for the frequency-tunable transmon in the coupler is replaced 
by a single Josephson junction for simplicity. 
Figure~\ref{fig-coupler}(b) shows the $ZZ$-coupling strength, $\zeta_{ZZ}$, of the coupler 
with typical parameter values~\cite{Collodo2020a}. 
Note that $ZZ$-coupling vanishing points [the white region in Fig.~\ref{fig-coupler}(b)] exist only in the straddling regime
[the dashed yellow box in Fig.~\ref{fig-coupler}(b)]. 
Thus, the single-transmon coupler cannot realize zero $ZZ$ coupling 
for highly detuned qubits. 
This is an essential contrast to the proposed double-transmon coupler.

\end{appendix}


\begin{thebibliography}{99}

\bibitem{Deutsch1995a}
D. Deutsch, A. Barenco, and A. Ekert, 
\textit{Universality in quantum computation}, 
Proc. R. Soc. Lond. A \textbf{449}, 669--677 (1995).

\bibitem{Barenco1995a}
A. Barenco, 
\textit{A universal two-bit gate for quantum computation}, 
Proc. R. Soc. Lond. A \textbf{449}, 679--683 (1995).


\bibitem{DiVincenzo1995a}
D. P. DiVincenzo, 
\textit{Two-bit gates are universal for quantum computation}, 
Phys. Rev. A \textbf{51}, 1015--1022 (1995).

\bibitem{Sleator1995a}
T. Sleator and H. Weinfurter, 
\textit{Realizable Universal Quantum Logic Gates}, 
Phys. Rev. Lett. \textbf{74}, 4087--4090 (1995).

\bibitem{Lloyd1995a}
S. Lloyd, 
\textit{Almost Any Quantum Logic Gate is Universal}, 
Phys. Rev. Lett. \textbf{75}, 346--390 (1995).

\bibitem{Barenco1995b}
A. Barenco, C. H. Bennett, R. Cleve, D. P. DiVincenzo, N. Margolus, P. Shor, T. Sleator, J. A.Smolin, and H. Weinfurter, 
\textit{Elementary gates for quantum computation}, 
Phys. Rev. A \textbf{52}, 3457--3467 (1995).

\bibitem{Nielsen}
M. A. Nielsen and I. L. Chuang, 
\textit{Quantum Computation and Quantum Information} (Cambridge University Press, 2000).

\bibitem{Raussendorf2007a}
R. Raussendorf and J. Harrington, 
\textit{Fault-Tolerant Quantum Computation with High Threshold in Two Dimensions}, 
Phys. Rev. Lett. \textbf{98}, 190504 (2007).

\bibitem{Raussendorf2007b}
R. Raussendorf, J. Harrington, and K. Goyal, 
\textit{Topological fault-tolerance in cluster state quantum computation}, 
New J. Phys. \textbf{9}, 199 (2007).

\bibitem{Fowler2009a}
A. G. Fowler, A. M. Stephens, and P. Groszkowski, 
\textit{High-threshold universal quantum computation on the surface code}, 
Phys. Rev. A \textbf{80}, 052312 (2009).

\bibitem{Wang2011a}
D. S. Wang, A. G. Fowler, and L. C. L. Hollenberg, 
\textit{Surface code quantum computing with error rates over 1\%}, 
Phys. Rev. A \textbf{83}, 020302(R) (2011).

\bibitem{Fowler2012a}
A. G. Fowler, M. Mariantoni, J. M. Martinis, and A. N. Cleland, 
\textit{Surface codes: Towards practical large-scale quantum computation}, 
Phys. Rev. A \textbf{86}, 032324 (2012).


\bibitem{Ballance2016a}
C. J. Ballance, T. P. Harty, N. M. Linke, M. A. Sepiol, and D. M. Lucas, 
\textit{High-Fidelity Quantum Logic Gates Using Trapped-Ion Hyperfine Qubits}, 
Phys. Rev. Lett. \textbf{117}, 060504 (2016).

\bibitem{Gaebler2016a}
J. P. Gaebler, T. R. Tan, Y. Lin, Y. Wan, R. Bowler, A. C. Keith, S. Glancy, K. Coakley, E. Knill, D. Leibfried, and D. J. Wineland, 
\textit{High-Fidelity Universal Gate Set for $^9$Be$^+$ Ion Qubits}, 
Phys. Rev. Lett. \textbf{117}, 060505 (2016).

\bibitem{Schafer2018a}
V. M. Sch\"{a}fer, C. J. Ballance, K. Thirumalai, L. J. Stephenson, T. G. Ballance, A. M. Steane, and D. M. Lucas, 
\textit{Fast quantum logic gates with trapped-ion qubits}, 
Nature \textbf{555}, 75--78 (2018).

\bibitem{Pino2021a}
J. M. Pino, J. M. Dreiling, C. Figgatt, J. P. Gaebler, S. A. Moses, M. S. Allman, C. H. Baldwin, 
M. Foss-Feig, D. Hayes, K. Mayer, C. Ryan-Anderson, and B. Neyenhuis, 
\textit{Demonstration of the trapped-ion quantum CCD computer architecture}, 
Nature \textbf{592}, 209--213 (2021).

\bibitem{Egan2021a}
L. Egan, D. M. Debroy, C. Noel, A. Risinger, D. Zhu, D. Biswas, M. Newman, M. Li, K. R. Brown, M. Cetina, and C. Monroe, 
\textit{Fault-tolerant control of an error-corrected qubit}, 
Nature \textbf{598}, 281--286 (2021).

\bibitem{Ryan2021a}
C. Ryan-Anderson, J. G. Bohnet, K. Lee, D. Gresh, A. Hankin, J. P. Gaebler, D. Francois, 
A. Chernoguzov, D. Lucchetti, N. C. Brown, T. M. Gatterman, S. K. Halit, K. Gilmore, 
J. A. Gerber, B. Neyenhuis, D. Hayes, and R. P. Stutz, 
\textit{Realization of Real-Time Fault-Tolerant Quantum Error Correction}, 
Phys. Rev. X \textbf{11}, 041058 (2021).


\bibitem{Barends2014a}
R. Barends, J. Kelly, A. Megrant, A. Veitia, D. Sank, E. Jeffrey, T. C. White, J. Mutus, 
A. G. Fowler, B. Campbell, Y. Chen, Z. Chen, B. Chiaro, A. Dunsworth, C. Neill, P. O'Malley, 
P. Roushan, A. Vainsencher, J. Wenner, A. N. Korotkov, A. N. Cleland, and J. M. Martinis, 
\textit{Superconducting quantum circuits at the surface code threshold for fault tolerance}, 
Nature \textbf{508}, 500--503 (2014).

\bibitem{Kelly2015a}
J. Kelly, R. Barends, A. G. Fowler, A. Megrant, E. Jeffrey, T. C. White, 
D. Sank, J. Y. Mutus, B. Campbell, Y. Chen, Z. Chen, B. Chiaro, A. Dunsworth, 
I.-C. Hoi, C. Neill, P. J. J. O'Malley, C. Quintana, P. Roushan, A. Vainsencher, J. Wenner, A. N. Cleland, and J. M. Martinis, 
\textit{State preservation by repetitive error detection in a superconducting quantum circuit}, 
Nature \textbf{519}, 66--69 (2015).

\bibitem{Arute2019a}
F. Arute, K. Arya, R. Babbush, D. Bacon, J. C. Bardin, R. Barends, R. Biswas, 
S. Boixo, F. G. S. L. Brandao, D. A. Buell, B. Burkett, Y. Chen, Z. Chen, B. Chiaro, 
R. Collins, W. Courtney, A. Dunsworth, E. Farhi, B. Foxen, A. Fowler, C. Gidney, 
M. Giustina, R. Graff, K. Guerin, S. Habegger, M. P. Harrigan, M. J. Hartmann, 
A. Ho, M. Hoffmann, T. Huang, T. S. Humble, S. V. Isakov, E. Jeffrey, Z. Jiang, 
D. Kafri, K. Kechedzhi, J. Kelly, P. V. Klimov, S. Knysh, A. Korotkov, F. Kostritsa, 
D. Landhuis, M. Lindmark, E. Lucero, D. Lyakh, S. Mandr\`{a}, J. R. McClean, M. McEwen, 
A. Megrant, X. Mi, K. Michielsen, M. Mohseni, J. Mutus, O. Naaman, M. Neeley, 
C. Neill, M. Y. Niu, E. Ostby, A. Petukhov, J. C. Platt, C. Quintana, E. G. Rieffel, 
P. Roushan, N. C. Rubin, D. Sank, K. J. Satzinger, V. Smelyanskiy, K. J. Sung, 
M. D. Trevithick, A. Vainsencher, B. Villalonga, T. Z. J. Yao, P. Yeh, A. Zalcman, 
H. Neven, and J. M. Martinis, 
\textit{Quantum supremacy using a programmable superconducting processor}, 
Nature \textbf{574}, 505--510 (2019).

\bibitem{Foxen2020a}
B. Foxen , C. Neill, A. Dunsworth, P. Roushan, B. Chiaro, A. Megrant, J. Kelly, 
Z. Chen, K. Satzinger, R. Barends, F. Arute, K. Arya, R. Babbush, D. Bacon, 
J. C. Bardin, S. Boixo, D. Buell, B. Burkett, Y. Chen, R. Collins, E. Farhi, A. Fowler, 
C. Gidney, M. Giustina, R. Graff, M. Harrigan, T. Huang, S. V. Isakov, E. Jeffrey, 
Z. Jiang, D. Kafri, K. Kechedzhi, P. Klimov, A. Korotkov, F. Kostritsa, D. Landhuis, 
E. Lucero, J. McClean, M. McEwen, X. Mi, M. Mohseni, J. Y. Mutus, O. Naaman, M. Neeley, 
M. Niu, A. Petukhov, C. Quintana, N. Rubin, D. Sank, V. Smelyanskiy, A. Vainsencher, 
T. C. White, Z. Yao, P. Yeh, A. Zalcman, H. Neven, and J. M. Martinis, 
\textit{Demonstrating a Continuous Set of Two-Qubit Gates for Near-Term Quantum Algorithms}, 
Phys. Rev. Lett. \textbf{125}, 120504 (2020).

\bibitem{Xu2020a}
Y. Xu, J. Chu, J. Yuan, J. Qiu, Y. Zhou, L. Zhang, X. Tan, Y. Yu, S. Liu, J. Li, F. Yan, and D. Yu, 
\textit{High-Fidelity, High-Scalability Two-Qubit Gate Scheme for Superconducting Qubits}, 
Phys. Rev. Lett. \textbf{125}, 240503 (2020).


\bibitem{Wu2021a}
Y. Wu, W.-S. Bao, S. Cao, F. Chen, M.-C. Chen, X. Chen, T.-H. Chung, H. Deng, Y. Du, 
D. Fan, M. Gong, C. Guo, C. Guo, S. Guo, L. Han, L. Hong, H.-L. Huang, Y.-H. Huo, 
L. Li, N. Li, S. Li, Y. Li, F. Liang, C. Lin, J. Lin, H. Qian, D. Qiao, H. Rong, H. Su, 
L. Sun, L. Wang, S. Wang, D. Wu, Y. Xu, K. Yan, W. Yang, Y. Yang, Y. Ye, J. Yin, 
C. Ying, J. Yu, C. Zha, C. Zhang, H. Zhang, K. Zhang, Y. Zhang, H. Zhao, 
Y. Zhao, L. Zhou, Q. Zhu, C.-Y. Lu, C.-Z. Peng, X. Zhu, and J.-W. Pan, 
\textit{Strong Quantum Computational Advantage Using a Superconducting Quantum Processor}, 
Phys. Rev. Lett. \textbf{127}, 180501 (2021).

\bibitem{Ficheux2021a}
Q. Ficheux, L. B. Nguyen, A. Somoroff, H. Xiong, K. N. Nesterov, M. G. Vavilov, and V. E. Manucharyan, 
\textit{Fast Logic with Slow Qubits: Microwave-Activated Controlled-Z Gate on Low-Frequency Fluxoniums}, 
Phys. Rev. X \textbf{11}, 021026 (2021).

\bibitem{Negirneac2021a}
V. Neg\^{i}rneac, H. Ali, N. Muthusubramanian, F. Battistel, R. Sagastizabal, M. S. Moreira, J.?F. Marques, 
W.?J. Vlothuizen, M. Beekman, C. Zachariadis, N. Haider, A. Bruno, and L. DiCarlo, 
\textit{High-Fidelity Controlled-Z Gate with Maximal Intermediate Leakage Operating at the Speed Limit in a Superconducting Quantum Processor}, 
Phys. Rev. Lett. \textbf{126}, 220502 (2021).

\bibitem{Sung2021a}
Y. Sung, L. Ding, J. Braum\"{u}ller, A. Veps\"{a}l\"{a}inen, B. Kannan, M. Kjaergaard, A. Greene, 
G. O. Samach, C. McNally, D. Kim, A. Melville, B. M. Niedzielski, M. E. Schwartz, J. L. Yoder, T. P. Orlando, S. Gustavsson, and W. D. Oliver, 
\textit{Realization of High-Fidelity CZ and $ZZ$-Free iSWAP Gates with a Tunable Coupler}, 
Phys. Rev. X \textbf{11}, 021058 (2021).

\bibitem{Stehlik2021a}
J. Stehlik, D. M. Zajac, D. L. Underwood, T. Phung, J. Blair, S. Carnevale, D. Klaus, 
G. A. Keefe, A. Carniol, M. Kumph, M. Steffen, and O. E. Dial, 
\textit{Tunable Coupling Architecture for Fixed-Frequency Transmon Superconducting Qubits}, 
Phys. Rev. Lett. \textbf{127}, 080505 (2021).

\bibitem{Sete2021a}
E. A. Sete, N. Didier, A. Q. Chen, S. Kulshreshtha, R. Manenti, and S. Poletto, 
\textit{Parametric-Resonance Entangling Gates with a Tunable Coupler}, 
Phys. Rev. Appl. \textbf{16}, 024050 (2021).

\bibitem{Kandala2021a}
A. Kandala, K. X. Wei, S. Srinivasan, E. Magesan, S. Carnevale, G. A. Keefe, D. Klaus, O. Dial, and D. C. McKay, 
\textit{Demonstration of a High-Fidelity CNOT Gate for Fixed-Frequency Transmons with Engineered $ZZ$ Suppression}, 
Phys. Rev. Lett. \textbf{127}, 130501 (2021).

\bibitem{Mitchell2021a}
B. K. Mitchell, R. K. Naik, A. Morvan, A. Hashim, J. M. Kreikebaum, B. Marinelli, W. Lavrijsen, K. Nowrouzi, D. I. Santiago, and I. Siddiqi, 
\textit{Hardware-Efficient Microwave-Activated Tunable Coupling between Superconducting Qubits}, 
Phys. Rev. Lett. \textbf{127}, 200502 (2021).

\bibitem{Ye2021a}
Y. Ye, S. Cao, Y. Wu, X. Chen, Q. Zhu, S. Li, F. Chen, M. Gong, C. Zha, H.-L. Huang, 
Y. Zhao, S. Wang, S. Guo, H. Qian, F. Liang, J. Lin, Y. Xu, C. Guo, L. Sun, N. Li, H. Deng, X. Zhu, and J.-W. Pan, 
\textit{Realization of high-fidelity CZ gates in extensible superconducting qubits design with a tunable coupler}, 
arXiv:2109.05680.



\bibitem{Noiri2022a}
A. Noiri, K. Takeda, T. Nakajima, T. Kobayashi, A. Sammak, G. Scappucci, and S. Tarucha, 
\textit{Fast universal quantum gate above the fault-tolerance threshold in silicon}, 
Nature \textbf{601}, 338--342 (2022).

\bibitem{Xue2022a}
X. Xue, M. Russ, N. Samkharadze, B. Undseth, A. Sammak, G. Scappucci, and L. M. K. Vandersypen, 
\textit{Quantum logic with spin qubits crossing the surface code threshold}, 
Nature \textbf{601}, 343--347 (2022).


\bibitem{Mills2022a} A. R. Mills, Charles R. Guinn, M. J. Gullans, A. J. Sigillito, M. M. Feldman, E. Nielsen, and J. R. Petta, 
\textit{Two-qubit silicon quantum processor with operation fidelity exceeding 99\%}, 
Sci. Adv. \textbf{8}, eabn5130 (2022).

\bibitem{Madzik2022a}
M. T. M\c{a}dzik, S. Asaad, A. Youssry, B. Joecker, K. M. Rudinger, E. Nielsen, K. C. Young, 
T. J. Proctor, A. D. Baczewski, A. Laucht, V. Schmitt, F. E. Hudson, K. M. Itoh, 
A. M. Jakob, B. C. Johnson, D. N. Jamieson, A. S. Dzurak, C. Ferrie, R. Blume-Kohout, and A. Morello, 
\textit{Precision tomography of a three-qubit donor quantum processor in silicon}, 
Nature \textbf{601}, 348--353 (2022).



\bibitem{Gong2021a}
M. Gong, S. Wang, C. Zha, M.-C. Chen, H.-L. Huang, Y. Wu, Q. Zhu, Y. Zhao, 
S. Li, S. Guo, H. Qian, Y. Ye, F. Chen, C. Ying, J. Yu, D. Fan, D. Wu, 
H. Su, H. Deng, H. Rong, K. Zhang, S. Cao, J. Lin, Y. Xu, L. Sun, C. Guo, N. Li, 
F. Liang, V. M. Bastidas, K. Nemoto, W. J. Munro, Y.-H. Huo, C.-Y. Lu, C.-Z. Peng, X. Zhu, and J.-W. Pan, 
\textit{Quantum walks on a programmable two-dimensional 62-qubit superconducting processor}, 
Science \textbf{372}, 948--952 (2021).




\bibitem{Yost2020a}
D. R. W. Yost, M. E. Schwartz, J. Mallek, D. Rosenberg, C. Stull, 
J. L. Yoder, G. Calusine, M. Cook, R. Das, A. L. Day,E. B. Golden, 
D. K. Kim, A. Melville, B. M. Niedzielski, W. Woods, A. J. Kerman, and W. D. Oliver, 
\textit{Solid-state qubits integrated with superconducting through-silicon vias}, 
npj Quant. Inf. \textbf{6}, 59 (2020).

\bibitem{Mallek2021a}
J. L. Mallek, D.-R. W. Yost, D. Rosenberg, J. L. Yoder, G. Calusine, 
M. Cook, R. Das, A. Day, E. Golden, D. K. Kim, J. Knecht, B. M. Niedzielski, M. Schwartz, 
A. Sevi, C. Stull, W. Woods, A. J. Kerman, and W. D. Oliver, 
\textit{Fabrication of superconducting through-silicon vias}, 
arXiv:2103.08536.

\bibitem{Kosen2021a}
S. Kosen, H.-X. Li, M. Rommel, D. Shiri, C. Warren, L. Gr\"{o}nberg, J. Salonen, 
T. Abad, J. Bizn\'{a}rov\'{a}, M. Caputo, L. Chen, K. Grigoras, G. Johansson, A. F. Kockum, 
C. Kri\v{z}an, D. P. Lozano, G. Norris, A. Osman, J. Fern\'{a}ndez-Pend\'{a}s, A. F. Roudsari, 
G. Tancredi, A. Wallraff, C. Eichler, J. Govenius, and J. Bylander, 
\textit{Building Blocks of a Flip-Chip Integrated Superconducting Quantum Processor}, 
arXiv:2112.02717.





\bibitem{Mundada2019a}
P. Mundada, G. Zhang, T. Hazard, and A. Houck, 
\textit{Suppression of Qubit Crosstalk in a Tunable Coupling Superconducting Circuit}, 
Phys. Rev. Appl. \textbf{12}, 054023 (2019).

\bibitem{Li2020a}
X. Li, T. Cai, H. Yan, Z. Wang, X. Pan, Y. Ma, W. Cai, J. Han, Z. Hua, X. Han, Y. Wu, 
H. Zhang, H. Wang, Y. Song, L. Duan, and L. Sun, 
\textit{Tunable Coupler for Realizing a Controlled-Phase Gate with Dynamically Decoupled Regime in a Superconducting Circuit}, 
Phys. Rev. Appl. \textbf{14}, 024070 (2020).

\bibitem{Collodo2020a}
M. C. Collodo, J. Herrmann, N. Lacroix, C. K. Andersen, A. Remm, S. Lazar, J.-C. Besse, T. Walter, A. Wallraff, and C. Eichler, 
\textit{Implementation of Conditional Phase Gates Based on Tunable $ZZ$ Interactions}, 
Phys. Rev. Lett. \textbf{125}, 240502 (2020).

\bibitem{Ni2021a}
Z. Ni, S. Li, L. Zhang, J. Chu, J. Niu, T. Yan, X. Deng, L. Hu, J. Li, Y. Zhong, S. Liu, F. Yan, Y. Xu, and D. Yu, 
\textit{Scalable method for eliminating residual $ZZ$ interaction between superconducting qubits}, 
arXiv:2111.13292.



\bibitem{Petrescu2021a}
A. Petrescu,  C. L. Calonnec, C. Leroux, A. D. Paolo, P. Mundada, S. Sussman, A Vrajitoarea, A. A. Houck, and A. Blais, 
\textit{Accurate methods for the analysis of strong-drive effects in parametric gates},
arXiv:2107.02343.


\bibitem{Jin2021a}
L. Jin, 
\textit{Implementing High-fidelity Two-Qubit Gates in Superconducting Coupler Architecture with Novel Parameter Regions},
arXiv:2105.13306.


\bibitem{Leroux2021a}
C. Leroux, A. D. Paolo, and A. Blais,
\textit{Superconducting Coupler with Exponentially Large On:Off Ratio},
Phys. Rev. Appl. \textbf{16}, 064062 (2021).



\bibitem{Miyanaga2021a}
T. Miyanaga, A. Tomonaga, H. Ito, H. Mukai, and J. S. Tsai, 
\textit{Ultrastrong Tunable Coupler Between Superconducting LC Resonators},
Phys. Rev. Appl. \textbf{16}, 064041 (2021).


\bibitem{Krantz2019a}
P. Krantz, M. Kjaergaard, F. Yan, T. P. Orlando, S. Gustavsson, and W. D. Oliver, 
\textit{A quantum engineer's guide to superconducting qubits}, 
Appl. Phys. Rev. \textbf{6}, 021318 (2019).

\bibitem{Yan2018a}
F. Yan, P. Krantz, Y. Sung, M. Kjaergaard, D. L. Campbell, T. P. Orlando, S. Gustavsson, and W. D. Oliver, 
\textit{Tunable Coupling Scheme for Implementing High-Fidelity Two-Qubit Gates}, 
Phys. Rev. Appl. \textbf{10}, 054062 (2018).

\bibitem{Niskanen2006a}
A. O. Niskanen, Y. Nakamura, and J.-S. Tsai, 
\textit{Tunable coupling scheme for flux qubits at the optimal point}, 
Phys. Rev. B \textbf{73}, 094506 (2006).

\bibitem{Niskanen2007a}
A. O. Niskanen, K. Harrabi, F. Yoshihara, Y. Nakamura, S. Lloyd, and J. S. Tsai, 
\textit{Quantum Coherent Tunable Coupling of Superconducting Qubits}, 
Science \textbf{316}, 723--726 (2007).




\bibitem{Allman2014a}
M. S. Allman, J. D. Whittaker, M. Castellanos-Beltran, K. Cicak, F. da Silva, M. P. DeFeo, F. Lecocq, A. Sirois, J. D. Teufel, J. Aumentado, and R. W. Simmonds, 
\textit{Tunable Resonant and Nonresonant Interactions between a Phase Qubit and LC Resonator},
Phys. Rev. Lett. \textbf{112}, 123601 (2014).


\bibitem{Whittaker2014a}
J. D. Whittaker, F. C. S. da Silva, M. S. Allman, F. Lecocq, K. Cicak, A. J. Sirois, J. D. Teufel, J. Aumentado, and R. W. Simmonds,
\textit{Tunable-cavity QED with phase qubits},
Phys. Rev. B \textbf{90}, 024513 (2014).

\bibitem{Chen2014a}
Y. Chen, C. Neill, P. Roushan, N. Leung, M. Fang, R. Barends, J. Kelly, B. Campbell, Z. Chen, B. Chiaro, A. Dunsworth, E. Jeffrey, A. Megrant, 
J. Y. Mutus, P. J. J. O'Malley, C. M. Quintana, D. Sank, A. Vainsencher, J. Wenner, T. C. White, M. R. Geller, A. N. Cleland, and J. M. Martinis,
\textit{Qubit Architecture with High Coherence and Fast Tunable Coupling}, 
Phys. Rev. Lett. \textbf{113}, 220502 (2014).

\bibitem{Neill2018a}
C. Neill,  P. Roushan,  K. Kechedzhi, S. Boixo, S. V. Isakov, V. Smelyanskiy, R. Barends,
B. Burkett, Y. Chen, Z. Chen, B. Chiaro, A. Dunsworth, A. Fowler, B. Foxen, 
R. Graff, E. Jeffrey, J. Kelly, E. Lucero, A. Megrant, J. Mutus, M. Neeley, C. Quintana,
D. Sank, A. Vainsencher, J. Wenner, T. C. White, H. Neven, and J. M. Martinis, 
\textit{A blueprint for demonstrating quantum supremacy with superconducting qubits},
Science \textbf{360}, 195--199 (2018).


\bibitem{Zhao2020a}
P. Zhao, P. Xu, D. Lan, J. Chu, X. Tan, H. Yu, and Y. Yu, 
\textit{High-Contrast $ZZ$ Interaction Using Superconducting Qubits with Opposite-Sign Anharmonicity}, 
Phys. Rev. Lett. \textbf{125}, 200503 (2020).

\bibitem{Ku2020a}
J. Ku, X. Xu, M. Brink, D. C. McKay, J. B. Hertzberg, M. H. Ansari, and B. L. T. Plourde, 
\textit{Suppression of Unwanted $ZZ$ Interactions in a Hybrid Two-Qubit System}, 
Phys. Rev. Lett. \textbf{125}, 200504 (2020).

\bibitem{Noguchi2020a}
A. Noguchi, A. Osada, S. Masuda, S. Kono, K. Heya, S. P. Wolski, H. Takahashi, T. Sugiyama, D. Lachance-Quirion, and Y. Nakamura, 
\textit{Fast parametric two-qubit gates with suppressed residual interaction using the second-order nonlinearity of a cubic transmon}, 
Phys. Rev. A \textbf{102}, 062408 (2020).

\bibitem{Xu2021a}
X. Xu and M. H. Ansari, 
\textit{$ZZ$ Freedom in Two-Qubit Gates}, 
Phys. Rev. Appl. \textbf{15}, 064074 (2021).

\bibitem{Zhao2021a}
P. Zhao, D. Lan, P. Xu, G. Xue, M. Blank, X. Tan, H. Yu, and Y. Yu, 
\textit{Suppression of Static $ZZ$ Interaction in an All-Transmon Quantum Processor}, 
Phys. Rev. Appl. \textbf{16}, 024037 (2021).

\bibitem{Finck2021a}
A. D. K. Finck, S. Carnevale, D. Klaus, C. Scerbo , J. Blair, T. G. McConkey , C. Kurter,
A. Carniol , G. Keefe, M. Kumph , and O. E. Dial, 
\textit{Suppressed Crosstalk between Two-Junction Superconducting Qubits with Mode-Selective Exchange Coupling},
Phys. Rev. Appl. \textbf{16}, 054041 (2021).



\bibitem{Tripathi2021a}
V. Tripathi, H. Chen, M. Khezri K.-W. Yip, E. M. Levenson-Falk, and D. A. Lidar, 
\textit{Suppression of crosstalk in superconducting qubits using dynamical decoupling}, 
arXiv:2108.04530.


\bibitem{flux-qubit-coupler}
Recently, couplers with a flux qubit, instead of a frequency-tunable transmon, 
have been proposed~\cite{Petrescu2021a,Jin2021a}.
These use three or more Josephson junctions, 
like our double-transmon coupler.
However, they are rather similar to the single-transmon coupler, 
and notably different from the double-transmon coupler.




\bibitem{mechanism}
This qualitative and classical explanation under rough approximations 
describes how we found the concept of the double-transmon coupler.
While this will be helpful for intuitive understanding of the mechanism, 
we must investigate the coupler in a more accurate manner 
for evaluating its performance, which is presented later.



\bibitem{hbar}
$\hbar$ and ${\phi_0=\Phi_0/(2\pi)}$ are, respectively, the reduced Planck constant 
and the reduced flux quantum ($\Phi_0$ is the flux quantum).



\bibitem{constraint}
This holds when the loop self-inductance is negligible.



\bibitem{Koch2007a}
J. Koch, T. M. Yu, J. Gambetta, A. A. Houck, D. I. Schuster, J. Majer, A. Blais, 
M. H. Devoret, S. M. Girvin, and R. J. Schoelkopf, 
\textit{Charge-insensitive qubit design derived from the Cooper pair box},
Phys. Rev. A \textbf{76}, 042319 (2007).





\bibitem{bound}
We found this fact by numerical studies. 
Its theoretical explanation is desirable, but left for future work.





\bibitem{parameters}
We choose the parameters in Fig.~\ref{fig-ZZ}(a) (out of the straddling regime), because in this work, 
we are interested in highly detuned qubits. 
We also choose ${\omega_4/(2\pi)=8.5}$~GHz as a relatively low value among the values satisfying the zero $ZZ$ coupling.
(We avoid the lower-bound value (about 8.3~GHz), 
because this may be not robust against small errors in parameter values.)



\bibitem{Martinis2014a}
J. M. Martinis and M. R. Geller, 
\textit{Fast adiabatic qubit gates using only $\sigma_z$ control}, 
Phys. Rev. A \textbf{90}, 022307 (2014).






\bibitem{spectator}
When there are more qubits, gate operations on other qubits will affect the performance, 
in particular, the zero ZZ coupling. This is an important issue, but 
its simulations need massive computations. 
Therefore, the study on this issue is left for future work.




\bibitem{Place2021a}
A. P. M. Place, L. V. H. Rodgers, P. Mundada, B. M. Smitham, M. Fitzpatrick, Z. Leng, A. Premkumar, 
J. Bryon, A. Vrajitoarea, S. Sussman, G. Cheng, T. Madhavan, H. K. Babla, X. H. Le, Y. Gang, 
B. J?ck, A. Gyenis, N. Yao, R. J. Cava, N. P. de Leon, and A. A. Houck, 
\textit{New material platform for superconducting transmon qubits with coherence times exceeding 0.3 milliseconds}, 
Nature Commun. \textbf{12}, 1779 (2021).

\bibitem{Wang2022a}
C. Wang, X. Li, H. Xu, Z. Li, J. Wang, Z. Yang, Z. Mi, X. Liang, T. Su, C. Yang, G. Wang, W. Wang, Y. Li, 
M. Chen, C. Li, K. Linghu, J. Han, Y. Zhang, Y. Feng, Y. Song, T. Ma, J. Zhang, R. Wang, P. Zhao, 
W. Liu, G. Xue, Y. Jin, and H. Yu, 
\textit{Towards practical quantum computers: transmon qubit with a lifetime approaching 0.5 milliseconds}, 
npj Quant. Inf. \textbf{8}, 3 (2022).



\bibitem{Hertzberg2021a}
J. B. Hertzberg, E. J. Zhang, S. Rosenblatt, E. Magesan, J. A. Smolin, J.-B. Yau, 
V. P. Adiga, M. Sandberg, M. Brink, J. M. Chow, and J. S. Orcutt, 
\textit{Laser-annealing Josephson junctions for yielding scaled-up superconducting quantum processors}, 
npj Quant. Inf. \textbf{7}, 129 (2021).

\bibitem{Kueng2016a}
R. Kueng, D. M. Long, A. C. Doherty, and S. T. Flammia, 
\textit{Comparing Experiments to the Fault-Tolerance Threshold}, 
Phys. Rev. Lett. \textbf{117}, 170502 (2016).

\bibitem{Nielsen2002a}
M. A. Nielsen, 
\textit{A simple formula for the average gate fidelity of a quantum dynamical operation}, 
Phys. Lett. A \textbf{303}, 249--252 (2002).



\bibitem{comment-A}
This value was chosen by trial and error.


\bibitem{Blais2021a}
A. Blais, A. L. Grimsmo, S. M. Girvin, and A. Wallraff, 
\textit{Circuit quantum electrodynamics}, 
Rev. Mod. Phys. \textbf{93}, 025005 (2021).





\end{thebibliography}
\end{document}